\documentclass[12pt]{article}
\usepackage{mathtools,amssymb,mathrsfs,microtype,tikz,slashed, amsmath}
\usepackage{physics}
\usepackage{color}
\usepackage[linktocpage]{hyperref}
\hypersetup{colorlinks=true,citecolor=blue,linkcolor=blue, urlcolor=blue, breaklinks=true}
\usepackage{cite}
\usepackage{moresize}
\usepackage{url}
\usepackage{cite}
\usepackage{mathtools}

\newcommand\subsetsim{\mathrel{\substack{
  \textstyle\subset\\[-0.2ex]\textstyle\sim}}}

\makeatletter\@addtoreset{equation}{section}\makeatother

\def\IZ{{\mathbb{Z}}}

\renewcommand{\title}[1]{\vbox{\center\LARGE{#1}}\vspace{5mm}}
\renewcommand{\author}[1]{\vbox{\center\large#1}\vspace{5mm}}
\newcommand{\address}[1]{\vbox{\center\em#1}}

\begin{document}

\begin{titlepage}

\begin{center}

% \hfill \\
% \hfill \\
\vskip 0.5cm

\title{Axion Domain Walls, Small Instantons, and Non-Invertible Symmetry Breaking}

\author{
Clay Córdova$^{1}$, Sungwoo Hong$^{2}$,  and
Lian-Tao Wang$^{1}$
}

\address{${}^1$Department of Physics, Kadanoff Center for Theoretical Physics \& Enrico Fermi Institute, University of Chicago}

\address{${}^2$Department of Physics, KAIST, Daejeon, 34141, Korea}

\end{center}

%

% \vfill

\abstract{

Non-invertible global symmetry often predicts degeneracy in axion potentials and carries important information about the global form of the gauge group.  When these symmetries are spontaneously broken they can lead to the formation of stable axion domain wall networks which support topological degrees of freedom on their worldvolume.  Such non-invertible symmetries 
can be broken by embedding into appropriate larger UV gauge groups where small instanton contributions lift the vacuum degeneracy, and provide a possible solution to the domain wall problem. We explain these ideas in simple illustrative examples and then apply them to the Standard Model, whose gauge algebra and matter content are consistent with several possible global structures.  Each possible global structure leads to different selection rules on the axion couplings, and various UV completions of the Standard Model lead to more specific relations.  As a proof of principle, we also present an example of a UV embedding of the Standard Model which can solve the axion domain wall problem. The formation and annihilation of the long-lived axion domain walls can lead to observables, such as gravitational wave signals. Observing such signals, in combination with the axion coupling measurements, can provide valuable insight into the global structure of the Standard Model, as well as its UV completion. 
}

\vfill

September 11, 2023

\vfill

\end{titlepage}

\eject

\tableofcontents

\unitlength = .8mm

\setcounter{tocdepth}{3}

\section{Introduction}\label{introsec}

Many aspects of gauge theory are deeply linked with topology.  Indeed, while the short distance spectrum of gauge bosons and their interactions depend only on local information, non-perturbative features such as magnetic monopoles and instanton effects depend on the global form of the gauge group, in particular its fundamental group \cite{Aharony:2013hda, Witten:2000nv}.  
A specific highlight is recent results constraining the phase diagram of Yang-Mills theory \cite{Gaiotto:2017yup, Cordova:2019uob} using subtle anomalies that can only be accessed by careful understanding of the global structure of the gauge group and its resulting implications for the quantization of instanton number.  Such anomalies have also been applied with particle physics motivations in \cite{Garcia-Etxebarria:2018ajm, Davighi:2019rcd,Hsin:2020nts, Anber:2021upc}.  These connections between quantum field theory and topology form the core of the generalized global symmetry program where, for instance, the spectrum of heavy probe particles, i.e.\ line operators, are interpreted in terms of a higher symmetry \cite{Gaiotto:2014kfa}. (See \cite{Cordova:2022ruw, Brennan:2023mmt, Schafer-Nameki:2023jdn, Bhardwaj:2023kri, Shao:2023gho} for recent reviews.)

Turning our attention to the Standard Model (SM), the matter content is consistent with four possible global structures for the gauge group:
\begin{equation}\label{smquo}
   \frac{ SU(3)_C \times SU(2)_L \times U(1)_Y}{\Gamma}~, ~~~\Gamma = \mathbf{1}, \ \mathbb{Z}_2, \ \mathbb{Z}_3, \ \mathbb{Z}_{6}~.
\end{equation}
A choice of global structure is needed to specify the possible magnetic lines in the theory which may arise dynamically as heavy probe particles in UV completions of the SM \cite{Tong:2017oea}.  Hence, different global structures can point to qualitatively different UV completions, such as grand unification theories and symmetry-breaking patterns that give rise to the SM.   Any experimental probes of the global structure of the SM gauge group can thus be a key clue toward possible UV dynamics.  However, probing the SM with terrestrial experiments, we expect the effect of the global structure to be exponentially suppressed due to the need to have a non-trivial spacetime topology or direct access to magnetic monopoles. This motivates us to consider possible effects in beyond the SM scenarios, and probes in astrophysical and cosmological settings.

Axions provide a natural arena to investigate the global form of the gauge group and related topological effects.  Indeed, axions have a rich menu of higher symmetries \cite{Hidaka:2020iaz, Hidaka:2020izy, Brennan:2020ehu, Choi:2022fgx, Yokokura:2022alv, Brennan:2023kpw} which provide universal constraints on possible UV completions \cite{Brennan:2020ehu, Choi:2022fgx}.   In our analysis below, a key role will be played by non-invertible shift symmetries of the axion sector, which are symmetries that are not represented by unitary operators acting on Hilbert space.  These operators were constructed in \cite{Choi:2022jqy, Cordova:2022ieu}, generalizing results of \cite{Choi:2021kmx, Kaidi:2021xfk}, and have recently played a role in particle physics and model building \cite{Choi:2022jqy, Choi:2022rfe, Niro:2022ctq, Cordova:2022fhg, Cordova:2023ent, strongCP}. In the context of axion physics, a hallmark of these symmetries is that the associated axion domain walls carry nontrivial topological degrees of freedom which may have distinct phenomenological consequences.

An important point that we illustrate below is that the possible axion couplings to gauge fields are sensitive to the global structure of the gauge group. In the context of the SM, the general coupling between an axion and the gauge fields can be parameterized as 
\begin{equation}\label{smactintro}
    S\supset \frac{i\ell_{1}}{8\pi^{2}} \int \frac{a}{f} F_{1}\wedge F_{1}+\frac{i\ell_{2}}{8\pi^{2}} \int \frac{a}{f} \mathrm{tr}(F_{2}\wedge F_{2})+\frac{i\ell_{3}}{8\pi^{2}}\int \frac{a}{f} \mathrm{tr}\left(F_{3}\wedge F_{3}\right)~.
\end{equation}
Here $F_{1}$ is the field strength of the $U(1)_Y$ gauge field associated with hypercharge with an integral normalization of charges, while $F_{2}$ is the field strength of the $SU(2)_L,$ and $F_{3}$ is the field strength of the $SU(3)_C$.  With different $\Gamma$ in \eqref{smquo}, the $\ell_i$s above are required to satisfy a set of selection rules derived in \autoref{sec:sm} below. In addition, different UV completions of the SM lead to more specific quantization conditions which we illustrate in examples. Experimental tests of such relations can give strong hints of both the global structure of the SM and its possible UV completion. 

One of the immediate consequences of the selection rules of the axion couplings is that, generically, at least some of the $\ell_i$s will be larger than unity. This leads to the presence of degenerate vacua in the axion potential.  In the context of non-trivial global structure, such degeneracy can be understood in an elegant way using the associated discrete non-invertible symmetries \cite{Cordova:2022ieu}. It is well-known that, for the specific case of $\ell_3 > 1$, these degenerate minima lead to the formation of significant topological structures in which multiple domain walls can end on an axion string \cite{Sikivie:1982qv}.  In the cosmological context, these string-domain wall networks are stable, and if so, can quickly dominate the energy density of the universe and lead to unacceptable evolution \cite{Sikivie:1982qv,Chang:1998tb,Hiramatsu:2012gg}. This is known as the domain wall problem \cite{Zeldovich:1974uw}. 

The understanding of the degeneracy in terms of non-invertible symmetries also points to ways of lifting such degeneracies \cite{Cordova:2022ieu}. Indeed, UV completions of the low energy gauge group often break the non-invertible symmetries. This is also consistent with the expectation that in general non-invertible symmetries, like all global symmetries, are accidental (see e.g.\ \cite{Reece:2023czb} for a discussion in the context of axions). In this case, the degeneracy will be lifted by an exponentially small amount by loops of virtual monopoles or, equivalently, through small instanton contributions from the UV gauge group \cite{Fan:2021ntg, Cordova:2022ieu}.  Such small instanton contributions to axion potentials have also been studied in \cite{Holdom:1982ex,Holdom:1985vx,Flynn:1987rs,Poppitz:2002ac,Gherghetta:2020keg,Agrawal:2017ksf,Agrawal:2017evu,Csaki:2019vte}.  

Once the axion degeneracy is lifted, there will be an imbalance of the pressure on the domain wall, which leads to the eventual collapse of the network and annihilation of the walls. This approach of solving the domain wall problem is referred to as introducing a bias in the potential \cite{Vilenkin:1981zs,Sikivie:1982qv,Gelmini:1988sf,Larsson:1996sp}. In general, a bias can be parameterized as $  V_{\rm b} \sim \epsilon f^4$, where $f$ is the axion decay constant. As briefly reviewed in \autoref{app:axionDWcosmo}, for the domain walls to annihilate before they dominate the energy of the universe, we require $\epsilon \gtrsim (m_a/M_{\rm Pl})^2$, where $m_a$ is the axion mass. Hence, an exponentially small breaking in the vacuum degeneracy is enough to solve the domain wall problem. This smallness of the breaking is also critical in the context of the QCD axion. Indeed, such a bias in the axion potential is a new contribution, in addition to those from QCD strong dynamics, and hence can potentially spoil the Peccei-Quinn mechanism \cite{Peccei:1977hh,Peccei:1977ur} in which the QCD axion as a solution to the strong CP problem \cite{Ghigna:1992iv,Barr:1992qq,Kamionkowski:1992mf,Dine:1992vx,Holman:1992us,Dobrescu:1996jp,Hiramatsu:2012sc}. This sets a rough bound on the size of the breaking $\epsilon < \theta_{\rm exp} \times m_a^2 f^{-2}$, with $\theta_{\rm exp} < 10^{-10}$. In contrast to the introduction of an ad hoc tiny bias, the mechanism proposed in this paper provides a natural origin for its size. 

In this work, we study the lifting of the degeneracies in detail. We work out several examples of possible UV completions of the SM,  corresponding to different possible global structures of the SM, and present the resulting axion potentials. As a demonstration, we construct an example where $\ell_3=3$ where a UV embedding $(U(1)_{Y}\times SU(3)_C)/\mathbb{Z}_{3}  \subset SU(9)$,  completely lifts the axion degeneracy by the small instantons and non-invertible symmetry breaking. 

If an axion is discovered, its coupling to the SM gauge bosons can be measured. Any pattern observed in such couplings could give us hints about the UV completion of the SM \cite{Agrawal:2022lsp}. The presence of the axion domain wall can offer more information. The annihilation of such domain walls can potentially lead to observable gravitational wave signals \cite{Vilenkin:1981zs,Preskill:1991kd,Chang:1998tb,Gleiser:1998na,Hiramatsu:2010yz,Kawasaki:2011vv,Saikawa:2017hiv}. At the same time, the nature of the domain wall, in particular, its topological worldvolume degrees of freedom, and associated spectrum of anyons depend on the global structure of the SM. Observing such signals together with the axion coupling measurement can be a clue to the global structure of the SM, as well as good evidence for the mechanism of solving the domain wall problem discussed in this work. 

The rest of this paper is organized as follows. In \autoref{sec:toy}, we illustrate the main physics mechanisms discussed in this paper with an example of an axion-Yang-Mills theory with the gauge group of $SU(N)/\mathbb{Z}_{p}$  extending the discussion in \cite{Cordova:2022ieu}. In \autoref{sec:sm}, we present the constraints on the axion couplings for different global structures of the SM, and work out several examples of the relations implied by UV completing the SM into different grand unification theories. We also present an explicit example of lifting the degeneracies in the axion potential using small instantons and non-invertible symmetry breaking. We then conclude in \autoref{sec:conclusion}.

\section{Global Structure and Axion Domain Walls}
\label{sec:toy}

\subsection{Axions Coupled to $SU(N)/\mathbb{Z}_{p}$}

Let us review the coupling of an axion and a gauge field.  We denote the periodic axion field by $a$ and the field strength by $F$.  The gauge group $G$ is assumed for now to be both simple and simply connected (e.g.\ $G\cong SU(N)$).  The kinetic terms and relevant couplings are:
\begin{equation}\label{action}
   S= \frac{1}{2}\int da \wedge *da+\frac{1}{g^{2}}\int \mathrm{tr}(F\wedge *F)+ \frac{i \ell}{8 \pi^2} \int \frac{a}{f} {\rm tr} (F \wedge F)~, \hspace{.5in} \ell \in \IZ~.
\end{equation}
Here $f$ is the axion decay constant setting the periodicity:
\begin{equation}\label{aperiodsc}
    a \sim a +2\pi f~,
\end{equation}
while $\ell$ is an integral coupling constant, the so-called domain wall number. It is integral because the instanton number is quantized:
\begin{equation}\label{in}
    \frac{1}{8 \pi^2} \int {\rm tr} (F \wedge F) \in \mathbb{Z}~,
\end{equation}
and hence $\ell$ is quantized to respect the periodicity \eqref{aperiodsc}.  

For $\ell>1$ there are stable axion domain walls.  This can be seen by inspecting the axion potential.  In the approximation where instantons are neglected, the potential is flat leading to a continuous shift symmetry and a circle's worth of vacua for the axion field.  Instantons lead to a potential $V(a)$ on this field space and we can interpret the instanton number as a Fourier mode number.  Specifically, we have: 
\begin{equation}\label{potis}
    V(a)=\sum_{n\in \mathbb{N}} \alpha_{n} \cos\left(\frac{\ell n a}{f}\right)~.
\end{equation}
In a weakly coupled theory, $\alpha_{n}\propto \exp(-\frac{8\pi^{2}n}{g^{2}})$ includes the instanton action as well as other factors that depend on the matter content and in particular is in general complex. Meanwhile, in a strongly coupled theory, we expect that $\alpha_{n}\propto \Lambda^{4}$ where $\Lambda$ is a dynamically generated scale.  Our analysis holds in both scenarios.    We note two features:
\begin{itemize}
    \item The model has an exact (invertible) $\mathbb{Z}_{\ell}$ symmetry under which the axion shifts as $a\rightarrow a+\frac{2\pi f}{\ell}$.  This is reflected in the fact that \eqref{potis} only includes Fourier modes that are multiples of $\ell$.
    \item A choice of minimum for the axion field spontaneously breaks this $\mathbb{Z}_{\ell}$ symmetry.  This leads to stable domain walls interpolating between the exact degenerate vacua.  
\end{itemize}

Let us now describe how this analysis depends on the global form of the gauge group.  For concreteness, we will focus on the case where at the Lie algebra level the group is isomorphic to $SU(N)$.  In this case, different global forms of $G$ are possible depending on the matter content.  Specifically, we consider the center $\mathbb{Z}_{N}$ of $SU(N)$ which acts by $N$-th roots of unity on the fundamental representation.  If there are matter fields for which this $\mathbb{Z}_{N}$ acts faithfully (e.g.\ if there is fundamental matter) then the gauge group must be globally $SU(N)$ since the matter must be a representation of the gauge group, not just the Lie algebra.  By contrast, if the matter does not form a faithful representation of the center $\mathbb{Z}_{N}$ there is more than one option for the global form of the gauge group.  

To parameterize the options, let $\mathbb{Z}_{K} \subset \mathbb{Z}_{N}$ be the subgroup of the center which acts trivially on all matter fields.  Explicitly this means that the matter content includes only representations whose associated Young tableaus have multiples of $K$ boxes.  A familiar example with $K= N$ is the situation of adjoint matter.   Next let $\mathbb{Z}_{p}$ be any subgroup of $\mathbb{Z}_{K}$ (so in particular $p$ divides $K$, and $K$ divides $N$), then the possible gauge groups of the model are $SU(N)/\mathbb{Z}_{p}$.  It is also useful to express these choices in terms of the one-form global symmetry of the model.  In general, there is both electric one-form symmetry $\mathbb{Z}_{K/p}^{(1)}$ counting Wilson lines that cannot be screened by dynamical matter fields, as well as magnetic one-form symmetry $\mathbb{Z}_{p}^{(1)}$ counting 't Hooft lines that cannot be screened by dynamical magnetic monopoles:
\begin{equation}\label{oneformsym}
    G\cong SU(N)/\mathbb{Z}_{p} \longleftrightarrow \text{one-form symmetry:}~~ \mathbb{Z}_{K/p}^{(1)}\times \mathbb{Z}_{p}^{(1)}~.
\end{equation}
We emphasize that a given model has a particular fixed gauge group.

We now return to our analysis of axion-Yang-Mills.  The reason the global form of the gauge group can enter our discussion is that the quantization of the instanton density is sensitive to this information. To describe this, we first recall that the possible gauge bundles of $SU(N)/\mathbb{Z}_{p}$ are richer than those of $SU(N)$.  Specifically, such bundles have a discrete magnetic flux specified by a degree two cohomology class $\omega$ (often referred to as a second Steifel-Whitney class)
\begin{equation}\label{omegadef}
    \omega \in H^{2}(M, \mathbb{Z}_{p})~,
\end{equation}
where above $M$ is the spacetime manifold. More concretely, this means that $\omega$ is an object which may be integrated over any two-cycle $\Sigma$ in spacetime yielding an integer which is well-defined modulo $p$:
\begin{equation}
    \int_{\Sigma} \omega \in \mathbb{Z}/p\mathbb{Z}\cong \mathbb{Z}_{p}~.
\end{equation}
At a pragmatic level, $SU(N)/\mathbb{Z}_{p}$ gauge theory differs from  $SU(N)$ gauge theory because we must sum over bundles with different values of $\omega.$  This sum can be non-trivial when spacetime has a sufficiently rich topology, or alternatively, in the presence of 't Hooft lines (heavy magnetic charges) which create two-cycles around which $\omega$ is non-vanishing. When $\omega$ is non-trivial the instanton number \eqref{in} is in general fractional, with a fractional part controlled by $\omega:$ 
\begin{equation}\label{fracform}
    \frac{1}{8\pi^{2}}\int_{M} {\rm tr} (F \wedge F)=\frac{N(N-1)}{2p^{2}}\mathcal{P}(\omega)\ \mod 1~.
\end{equation}
Above, the notation mod $1$ means that the difference between the left and right-hand sides is an integer. Meanwhile, $\mathcal{P}(\omega)$ denotes the Pontryagin square cohomology operation. For the purposes of our calculations below, we can interpret $\mathcal{P}(\omega)$ as simply\footnote{More precisely, $\mathcal{P}$ is a cohomology operation of type $H^{2}(M,\mathbb{Z}_{p})\rightarrow H^{4}(M,\mathbb{Z}_{2p})$ for even $p$ while for odd $p,$ $\mathcal{P}(\omega)=\omega \cup \omega \in H^{4}(M,\mathbb{Z}_{p}).$ Note that on spin manifolds, which is our concern here, $\mathcal{P}(\omega)$ is always even for $p$ even, while for $p$ odd $2$ is invertible (and hence can be moved to the numerator).   In practice for the calculations carried out below, one can often treat $\mathcal{P}(\omega)$ by picking lifts as simply $\int_{M} \omega ^{2}$ with the appropriate coefficients.  The evaluation of $\mathcal{P}(\omega)$ when no such lift exists is more subtle but standard.}
\begin{equation}\label{liftform}
    \mathcal{P}(\omega)=\int_{M}\omega^{2} \in \begin{cases}
        \mathbb{Z}_{2p}~, & p~\rm{even}~,\\
        \mathbb{Z}_{p}~, & p~\rm{odd}~.
    \end{cases} 
\end{equation}
  Note that $p$ divides $N$, and moreover, $2$ divides $\mathcal{P}(\omega),$ hence the minimum non-trivial value of the right-hand side of \eqref{fracform} is $1/p.$  An instanton of this type is referred to as fractional.  

The existence of fractional instantons means that for $SU(N)/\mathbb{Z}_{p}$ the allowed values of the coupling constant $\ell$ are modified.   This is because the axion periodicity is a gauge redundancy on the field space and hence the action must be exactly invariant under shifts of the axion by its period.  Inspecting \eqref{fracform} and \eqref{action}, we see that for an axion coupled to an $SU(N)/\mathbb{Z}_{p}$ the allowed values are now:
\begin{equation}\label{periodp}
    \ell\in p \mathbb{Z}~.
\end{equation}
Thus, the minimum allowed coupling is larger to account for the possibility of fractional instantons.\footnote{Equivalently, one may also say that the periodicity of the axion is larger by a factor of $p$. We chose this normalization convention for uniformity across different choices of global form for the gauge group.}  

Let us also inspect the axion potential for an axion coupled to $SU(N)/\mathbb{Z}_{p}$. To account for fractional instantons, the general Fourier expansion now takes the form
\begin{equation}\label{potp}
     V(a)=\sum_{n\in \mathbb{N}} \alpha_{n} \cos\left(\frac{\ell n a}{pf}\right)~.
\end{equation}
Here, $n/p$ is the instanton number as measured by the left-hand side of \eqref{fracform}, and hence for $n$ not a multiple of $p$ the associated Fourier modes of the potential would come from factional instantons.  However, although such fractional instantons generally contribute to the partition function, the axion potential calculation does not have sufficient topology (or equivalently magnetic charges) to activate such configurations. Thus, the potential only gets contributions from integrally quantized instantons.  Hence in fact $\alpha_{n}$ in \eqref{potp} vanishes unless $n$ is divisible by $p$ and the potential takes the same form as \eqref{potis}.  We make the following observations:
\begin{itemize}
    \item The Fourier modes corresponding to fractional instantons are missing in the potential.  As a result the axion potential again has $\ell$ distinct minima and is invariant under shifts $a \rightarrow a+\frac{2\pi  f}{\ell}~.$  In particular this means that even for the smallest allowed value $\ell=p$, there are multiple minima of the potential.
    
    \item The presence of multiple minima reflects a symmetry of the model.  Some of the symmetries are the familiar invertible axion shifts $a \rightarrow a+\frac{2\pi f}{\ell/p}$ that also arose above in our analysis of $p=1$. To emphasize the invertiblility of these symmetries we denote this group as $\mathbb{Z}_{\ell/p}^{I}$.  There are also symmetries of the model that achieve finer shifts of the axion, $a \rightarrow a+\frac{2\pi f}{\ell}$. In general, these are non-invertible symmetries and we indicate them by $\mathbb{Z}_{\ell}^{NI}.$  They enforce the vanishing of the Fourier modes in the potential which correspond to fractional instantons. Note that the invertible symmetries are a subgroup of the non-invertible symmetries $\mathbb{Z}_{\ell/p}^{I} \subseteq \mathbb{Z}_{\ell}^{NI}.$

    \item A choice of minimum for the axion field spontaneously breaks the $\mathbb{Z}_{\ell}^{NI}$ non-invertible symmetry.  This leads to stable domain walls interpolating between the exact degenerate vacua.  
\end{itemize}

\subsection{Non-Invertible Symmetry and Domain Wall Physics}
\label{sec:theWall}

Let us elaborate on the non-invertible symmetry and its physical consequences for the axion domain walls.  In this case, the signature of a domain wall being non-invertible is that it supports a non-trivial topological quantum field theory on its worldvolume that couples to the bulk gauge fields.  This construction was derived in \cite{Choi:2022jqy, Cordova:2022ieu} and we review it below.  

In general, the worldvolume theory on the defect is an abelian Chern-Simons gauge theory familiar from the physics of the fractional quantum Hall effect. Focusing on the case $SU(N)/\mathbb{Z}_{p}$, we consider the minimal domain wall $\mathcal{D}$.  Across the wall, the axion transforms as $a\rightarrow a+\frac{2\pi f}{\ell}$. More formally, we say that this $\mathbb{Z}_{\ell}^{NI}$ non-invertible symmetry transformation is generated by defect $\mathcal{D}$. 
On the wall locus (which we also indicate by $\mathcal{D}$) there is a new dynamical gauge field $b$ with action:
\begin{equation}\label{dwaction}
    S_{\mathcal{D}}=\frac{i p}{4\pi}\int_{\mathcal{D}} b \wedge db+i\int_{\mathcal{D}} b\wedge \omega~,
\end{equation}
where the first term above is the Chern-Simons action for this three-dimensional gauge field, and the second term indicates the coupling of the wall degrees of freedom to the bulk.  

Notice that there is no Yang-Mills kinetic term for the gauge field $b$, so this action is topological. In particular, when we ignore the coupling to the bulk, the topological degrees of freedom may be integrated out and effectively ignored since they do not carry energy or momentum.  The situation is more interesting when $\omega$ is non-trivial and the bulk and wall are coupled.  Here we recall that $\omega$ is the discrete magnetic flux of $SU(N)/\mathbb{Z}_{p}$.  This means for instance, that the integral of $\omega$ over a two-sphere surrounding a heavy monopole ('t Hooft line) of $SU(N)/\mathbb{Z}_{p}$ measures its conserved magnetic charge in $\mathbb{Z}_{p}.$  Inspecting the coupling between $b$ and $\omega$ in \eqref{dwaction}  we see that such field configurations activate the Wilson lines of $b$ in the domain wall gauge theory.  

To be even more explicit, we can consider an example where the domain wall has the spatial topology of a two-sphere extended in time separating an interior axion minima from an exterior axion minima.  Suppose moreover, that somewhere in the interior region is a heavy magnetic monopole whose origin may be some yet unknown UV gauge dynamics.  From the IR point of view, this is modeled as a magnetic 't Hooft line operator of the IR theory carrying magnetic charge $q$ defined modulo $p$.  Then on the wall, we find that there is an anyon carrying $b$ charge $q$ of fractional spin:
\begin{equation}
    h(q)=\frac{q^{2}}{2p}~, \hspace{.1in} \mod 1/2~.
\end{equation}
It is also clear that further fractional hall physics can be realized using these domain walls.  For instance, the edge modes of the wall $\mathcal{D}$ are chiral and can carry energy and momentum which may lead to distinct phenomenological signatures. 

As noted in \cite{Cordova:2022ieu} the non-invertibility of $\mathcal{D}$ also has a significant impact on the fusion algebra of the defects which leads to natural breaking mechanisms for these symmetries.  Specifically, let us consider colliding the wall $\mathcal{D}$ with the antiwall $\overline{\mathcal{D}}$.  For ordinary unitary symmetries, this leads to a trivial wall configuration since $\mathcal{D}$ is a unitary transformation with $\overline{\mathcal{D}}$ its inverse.  By contrast in the case of the non-invertible symmetry, we have the more exotic fusion rule:\footnote{Formula \eqref{fusion} is schematic, neglecting coefficients.  See \cite{Cordova:2022ieu} for a precise version. }
\begin{equation}\label{fusion}
    \mathcal{D}\times \mathcal{\overline{D}}\sim \sum_{\text{two-cycles} ~\Sigma \subset \mathcal{D}}\exp\left(\frac{2\pi i}{p}\int_{\Sigma} \omega\right)~.
\end{equation}
The right-hand side is a sum over topological surface operators defining the one-form magnetic symmetry of $\mathbb{Z}_{p}^{(1)}$ referred to in \eqref{oneformsym}.  

To make contact with symmetry breaking, we observe that the magnetic one-form symmetries are present because the dynamical particles in the theory cannot screen all possible probe magnetic charges.  Thus, if this model is viewed as the IR of a parent UV theory where such probe magnetic charges are promoted to dynamical monopoles arising via Higgsing, then in the UV the magnetic one-form symmetries will be broken. Concretely this means that the operators appearing on the right-hand side of \eqref{fusion} will cease to be topological defects.  Since the fusion of topological defects must always be topological, the fusion algebra \eqref{fusion} in turn implies that the domain wall $\mathcal{D}$ is also destabilized in any such UV model.  Physically this algebraic mechanism may be viewed as the generation of additional short distance contributions to the axion potential from loops of magnetic monopoles \cite{Fan:2021ntg}, or equivalently as we discuss below, from small instantons.

\subsection{Lifting Vacuum Degeneracy with Small Instantons}\label{sec:si}

As described in \autoref{introsec} and briefly reviewed in  \autoref{app:axionDWcosmo}, the presence of exactly stable domain walls can lead to problematic cosmology where networks of domain walls dominate the energy density of the universe.  For this reason, it is desirable to consider mechanisms that can split the degeneracy.  Here we discuss examples where the axion potential is modified by contributions from an ultraviolet gauge group $G_{\rm UV}$ which is Higgsed to $G_{\rm IR}$.  Our specific focus is on so-called small instantons, i.e.\ instantons of the ultraviolet group which are confined by the Higgsing process $G_{\rm UV}\rightarrow G_{\rm IR}.$  Depending on the nature of the embedding, the small instantons of $G_{\text{UV}}$ may appear from the IR point of view to be fractional instantons of $G_{\text{IR}},$ and hence may contribute missing Fourier modes in the axion potential needed to lift degeneracies between minima and hence eliminate domain walls.  In particular, this mechanism may in general break both invertible and non-invertible symmetries of the infrared theory. 

Here we focus on simple examples illustrating this point following \cite{Cordova:2022ieu}. (In particular for now we assume both $G_{\text{UV}}$ and $G_{\text{IR}}$ are simple.)  In later sections, we present examples more relevant to the Standard Model. Consider again the axion potential generated by the coupling to the infrared gauge group.  We wish to express this in terms of contributions from UV instantons.  To do so we require the knowledge of how the instanton numbers in the UV and IR are related.  This information is encoded by the so-called index of embedding $c \in \mathbb{N}$.  An infrared gauge field configuration with instanton number one has instanton number $c$ as seen from the point of view of the ultraviolet gauge group.  Concretely, one may compute the index of embedding $c$ using representation theory as follows.  Consider any representation $\mathbf{R}$ of $G_{\text{UV}}$.  Under the subgroup $G_{\text{IR}},$ $\mathbf{R}$ decomposes into a direct sum $\oplus_{i}\mathbf{R}_{i}.$  Then, the embedding index is the ratio of Dynkin indices of the representations:
\begin{equation}\label{index}
    c=\frac{\sum_{i}T(\mathbf{R}_{i})}{T(\mathbf{R})}~.
\end{equation}

Notice also that the index $c$ precisely tells us about the ratio of traces needed to define the action, i.e. ${\rm tr}_{\text{IR}}=c~{\rm tr}_{\text{UV}}$.  Therefore we have:
\begin{equation}
    \frac{\ell_{\text{IR}}}{8 \pi^2} \int \frac{a}{f} {\rm tr}_{\text{IR}} (F \wedge F)= \frac{c \ \ell_{\text{UV}}}{8 \pi^2} \int \frac{a}{f} {\rm tr}_{\text{UV}} (F \wedge F)~.
    \label{eq:aggUV}
\end{equation}
Therefore we obtain the following matching conditions on the $\ell_{i}$ and gauge couplings (up to threshold corrections):
\begin{equation}\label{lmatch}
    \ell_{\text{IR}}=c \ \ell_{\text{UV}}~, \hspace{.5in} \frac{1}{g^{2}_{\text{IR}}(v)}=\frac{c}{g^{2}_{\text{UV}}(v)}~,
\end{equation}
where above, $v$ is the Higgsing scale associated to the breaking $G_{\text{UV}}\rightarrow G_{\text{IR}}.$  

We can now compare the axion potential as computed from the infrared and ultraviolet and thereby elucidate the contributions of small instantons.  The infrared potential takes the form: 
\begin{equation}\label{virmatch}
    V_{\text{IR}}(a)=\sum_{n \in \mathbb{N}}\alpha_{n}\cos\left(\frac{\ell_{\text{IR}}na}{f}\right)~, \hspace{.5in}\alpha_{n}\propto \exp\left(-\frac{8\pi^{2}n}{g^{2}_{\text{IR}}}\right)~,
\end{equation}
while the ultraviolet potential takes the form: 
\begin{equation}
    V_{\text{UV}}(a)=\sum_{m\in \mathbb{N}}\beta_{m}\cos\left(\frac{\ell_{\text{UV}}m a}{f}\right)~, \hspace{.5in}\beta_{m}\propto \exp\left(-\frac{8\pi^{2}m}{g^{2}_{\text{UV}}}\right)~.
\end{equation}
Using \eqref{lmatch} we can rewrite the above as
\begin{equation}
    V_{\text{UV}}(a)=\sum_{m \in \mathbb{N}}\alpha_{m/c}\cos\left(\frac{\ell_{\text{IR}}ma}{cf}\right)~.
\end{equation}
Comparing to \eqref{virmatch}, we see that the UV contributions to the potential with mode number $m$ divisible by the embedding index $c$ reproduce the IR potential, while the remaining modes appear, from the IR point of view, as fractional contributions with effective instanton number $m/c.$  

It is these small instanton contributions to the axion potential that may qualitatively alter the axion minima and domain walls.  For instance, such terms can provide a naturally small splitting between nearly degenerate ground states.  Such a mechanism is particularly attractive when the axion couples to color to solve the strong CP problem, where violations of PQ symmetry must be exquisitely small.  We can estimate the size of this splitting as simply:
\begin{equation}
    \Delta(V) \sim v^4 \exp \left( - \frac{8 \pi^2 }{g^2_{\text{UV}}(v)}\right),
\end{equation}
where $v$ is scale of $G_{\rm UV} \to G_{\rm IR}$ Higgsing.

\subsection{Examples with Small Instantons}

Let us illustrate the above discussion with several examples. These examples are chosen for simplicity. Their global structures are different from that of the SM. At the same time, they could be phenomenologically relevant if they are realized in some dark sector. 

\subsubsection{Lifts of $SU(N)/\mathbb{Z}_{N}$}

Consider the case where the IR group is $SU(N)/\mathbb{Z}_{N}.$  This means that the allowed matter fields are neutral under the full center $\mathbb{Z}_{N}.$ One such representation is the adjoint $\mathbf{N^{2}-1}$. We consider a UV lift where this becomes the fundamental.  This can be achieved for $G_{\text{UV}}\cong SU(N^{2}-1)$.
\begin{equation}\label{subranchExample1}
    \begin{tabular}{cccc}
      & $SU(N^{2}-1)$  & $\rightarrow$ & $SU(N)/\mathbb{Z}_{N}$  \\
       \hline \\
{\rm Representation:}   &  $\mathbf{N^{2}-1}$  & $\rightarrow$ & $\mathbf{N^{2}-1}$
    \end{tabular}
\end{equation}
Note that this embedding of subgroups requires the $\mathbb{Z}_{N}$ quotient on the infrared group.  Indeed, for this embedding, the center of $SU(N)$ acts trivially on all representations of $SU(N^{2}-1)$.  There are of course $SU(N)$ subgroups of $SU(N^{2}-1)$ that does not require the quotient, but these do not achieve the branching rule \eqref{subranchExample1}.

A particular example of the above which appeared in \cite{Cordova:2022ieu} is the special case $N=2$.  In this case, the embedding in question is $SU(2)/\mathbb{Z}_{2}\cong SO(3) \subset SU(3)$, which can be understood as the restriction of the $SU(3)$ matrices to its real subgroup.  

It is straightforward to evaluate the index of embedding in this general example \eqref{subranchExample1}.  Using the definition \eqref{index}, we find that:
\begin{equation}
    c=2N~ \ \Longrightarrow \ \ell_{\rm IR}=2N \ell_{\rm UV}~.
\end{equation}
From these group theory considerations, we learn several facts about an axion coupled to $SU(N)/\mathbb{Z}_{N}$ which lifts in the UV to an axion coupled to $SU(N^{2}-1)$:
\begin{itemize}
    \item Any such model must have $\ell_{\text{IR}}$ divisible by $2N$.  Note that this is a stronger quantization condition than \eqref{periodp}.  Indeed, a specific choice of UV completion can impose additional requirements on the couplings $\ell_{\text{IR}}$.  Correspondingly, knowing the value of $\ell_{\rm IR}$ can be a clue towards possible UV physics.
    \item Since the embedding index $c>1$ the UV of such a toy model can solve IR domain wall problems.  Specifically, the IR appears to have $2N \ell_{\text{UV}}$ minima, whereas the UV has only $\ell_{\text{UV}}$ minima.  For instance, if $\ell_{\text{UV}}=1$,  all minima are lifted.  
    \item The lifting of vacua can be interpreted in terms of symmetry breaking. The IR model has a $\mathbb{Z}_{2N \ell_{\rm UV}}^{NI}$ symmetry with a $\mathbb{Z}_{2 \ell_{\rm UV}}^{I}$ invertible subgroup.  Meanwhile the UV has only a $\mathbb{Z}_{\ell_{\rm UV}}^{I}$ global symmetry. Thus the additional symmetries are broken by small instantons. 
\end{itemize}

\subsubsection{Lifts of $(SU(r)\times SU(s))/\mathbb{Z}_{\gcd(r,s)}$}\label{toy2}

Let us also give an example where the infrared is not a simple group.  This is also useful preparation for the case of the Standard Model discussed below.  We consider:
\begin{equation}
    G_{\text{IR}}\cong (SU(r)\times SU(s))/\mathbb{Z}_{\gcd(r,s)}~,
\end{equation}
where the quotient above is by the common center of the two factors.  The allowed matter fields are neutral under the common center, and hence a minimal representation is the bifundamental $(\mathbf{r},\mathbf{s}).$  We consider a lift where this becomes the fundamental of the UV group.  This can be achieved for $G_{UV}\cong SU(rs)$.
\begin{equation}\label{subranch2}
    \begin{tabular}{cccc}
      &  $SU(rs)$  & $\rightarrow$ & $(SU(r)\times SU(s))/\mathbb{Z}_{\gcd(r,s)}$  \\
       \hline \\
      {\rm Representation: } & $\mathbf{rs}$  & $\rightarrow$ & $(\mathbf{r},\mathbf{s})$
    \end{tabular}
\end{equation}
Again we note that this particular branching rule is only correct for the global structure on the IR subgroup identified above.  (There are other subgroups $SU(r)\times SU(s)$ of $SU(rs)$ that do not have the quotient and result in different branchings.)  As in the previous example, we can gain intuition from the special case of $r=s=2$.  In that case, the embedding in question results from restricting $SU(4)$ to its real subgroup $SO(4)\cong (SU(2)\times SU(2))/\mathbb{Z}_{2},$ and the branching is the familiar fact that the vector representation of $SO(4)$ is a bifundamental of $SU(2)$ spinors. 

To understand the relationship between the IR and UV axion couplings, we must slightly generalize our discussion to account for the fact that the IR group is not simple.  We express the relevant terms in the IR action as:
\begin{equation}
     S\supset \frac{i\ell_{r}}{8\pi^{2}} \int \frac{a}{f} \mathrm{tr}(F_{r}\wedge F_{r})+\frac{i\ell_{s}}{8\pi^{2}} \int \frac{a}{f} \mathrm{tr}(F_{s}\wedge F_{s})~,
\end{equation}
where $F_{j}$ for $j=r,s$ are the gauge fields for the (locally) $SU(j)$ gauge fields and $\ell_{j}$ the associated axion couplings. We aim to determine $\ell_{j}$ in terms of $\ell_{\rm UV}.$  This can be carried out using the branching \eqref{subranch2} to evaluate the associated indices of embedding of each of the factors of the IR group leading to the result:
\begin{equation}\label{srsex}
    \ell_{r}=s \ell_{\rm UV}~, \hspace{.2in} \ell_{s}=r\ell_{\rm UV}~.
\end{equation}

Note also that the IR axion potential in this model is double sum over instantons:
    \begin{equation}
    V_{\text{IR}}(a)=\sum_{n,m=0}^{\infty}\alpha_{n}\beta_{m}\cos\left(\frac{\ell_{r}n a}{f}+\frac{\ell_{s}m a}{f}\right)~.
\end{equation}
 Hence the IR has multiple minima whenever $\ell_{r}$ and $\ell_{s}$ have a common factor. In particular, if we inspect those examples that satisfy \eqref{srsex}, we see that in the case where the IR gauge group is not simply connected (i.e.\ $r$ and $s$ have a common factor), then there are multiple IR minima and associated domain walls.  However, these domain walls can be lifted by small instantons of $SU(rs)$.  For example, when $\ell_{\rm UV}=1$ the short distance potential has a unique minimum, and hence all IR degeneracies are broken.  As above, this can be interpreted in terms of non-invertible symmetry breaking.

\section{Axions Coupled to the Standard Model and Beyond}
\label{sec:sm}

In this section, we upgrade our analysis from the toy models of the previous section to the Standard Model.  The gauge group is:
\begin{equation}\label{smgg}
  (SU(3)_{C}\times SU(2)_{L} \times U(1)_{Y})/\mathbb{Z}_{p}~,
\end{equation}
where $p$ can be $1, 2, 3, 6.$  Our goal is to elucidate how the global structure of the gauge group (choice of $p$) influences the spectrum of axion domain walls and to investigate when non-invertible symmetries point to potential UV completions where small instantons can solve domain wall problems.  To this end, we first determine how the global structure constrains the coupling of axions to the Standard Model and then generalize to UV uplifts including grand unified models and other more exotic unification patterns. 

Below, we parameterize the relevant terms in the action as:
\begin{equation}\label{smact}
    S\supset \frac{i\ell_{1}}{8\pi^{2}} \int \frac{a}{f} F_{1}\wedge F_{1}+\frac{i\ell_{2}}{8\pi^{2}} \int \frac{a}{f} \mathrm{tr}(F_{2}\wedge F_{2})+\frac{i\ell_{3}}{8\pi^{2}}\int \frac{a}{f} \mathrm{tr}\left(F_{3}\wedge F_{3}\right)
\end{equation}
Here $F_{1}$ is the field strength of the (locally) $U(1)_{Y}$ gauge field associated to hypercharge with integrally quantized charges, while $F_{2}$ is the field strength of the (local) $SU(2)_{L},$ and $F_{3}$ is the field strength of the local $SU(3)_{C}$.  

\subsection{Quantization Conditions on Axion Couplings}
\label{sec:couplings}

We must first deduce the allowed values of the couplings $\ell_{i}$.  This can be straightforwardly carried out using the fractional instanton analysis described above together with knowledge of the spectrum of magnetic line operators.  Indeed, it is the existence of magnetic line operators in $SU(N)/\mathbb{Z}_{p}$ which gives rise to the $\mathbb{Z}_{p}$-valued discrete magnetic flux $\omega$ in \eqref{omegadef} that enters the formula \eqref{fracform} determining the pattern of fractional instantons. 

In generalizing to the Standard Model, we must know the generalization of these ideas to gauge group $U(1)$ and also to product groups.  Consider first a quotient that acts only on an abelian factor, so that we are considering gauge group $U(1)/\mathbb{Z}_{p}.$  This quotient means that the allowed charges of matter representations are multiples of $p.$  Then, Dirac quantization allows fractional magnetic charges quantized in units of $1/p.$  This is expressed in terms of the gauge field by stating that the fluxes obey:
\begin{equation}
    p \int_{\Sigma} \frac{F_{1}}{2\pi} \in \mathbb{Z}~,
\end{equation}
where $\Sigma$ is any two-cycle, e.g. a sphere surrounding a magnetic charge.  

To upgrade to the Standard Model we now only need to know how to incorporate quotients that act simultaneously on two or more groups.  For instance, we can consider the unitary group $U(N)$
\begin{equation}
    U(N)\cong (SU(N)\times U(1))/\mathbb{Z}_{N}~.
\end{equation}
The matter fields are neutral under $\mathbb{Z}_{N}$, but this still permits them to be charged under the center of each factor above individually.  For instance $(\mathbf{N},+1)$ is an allowed representation.  Correspondingly, fractional magnetic charges of both factors of the gauge group are permitted, but in order to satisfy Dirac quantization with $(\mathbf{N},+1)$, the fractional parts of the magnetic charge must be appropriately correlated.  Denoting the discrete magnetic flux of $SU(N)/\mathbb{Z}_{N}$ by $\omega(A_{N})$ the flux quantization condition is: 
\begin{equation}
     \int_{\Sigma}\frac{F_{1}}{2\pi}=\frac{1}{N}\int_{\Sigma}\omega(A_{N}) ~~, \mod 1~,
\end{equation}
where again the notation$\mod 1$ means that the difference between the left and right-hand sides is an integer.  Alternatively, we may write the same condition above as 
\begin{equation}
    \frac{F_{1}}{2\pi}=\frac{1}{N}\omega(A_{N})+X~,
\end{equation}
where $X\in H^{2}(M,\mathbb{Z})$ and each term above is viewed as a cohomology class (so that equality holds upon integration on any two-cycle in spacetime $M$.)  Below we often adopt this notation.

We now analyze each choice of $p$ in the Standard Model gauge group \eqref{smgg}.  As we will see each choice of $p$ implies distinct conditions on the $\ell_{i}$ and hence knowledge of the $\ell_{i}$ (e.g.\ from potential future experiments) can be a clue to the global structure of the gauge group.

\paragraph{$p=1$:} There is no quotient so all matter representations are allowed.  Correspondingly there are no fractional magnetic charges, and hence no fractional instantons.  This implies that the axion coupling quantization condition is simply: $\ell_{1}, \ell_{2}, \ell_{3} \in \mathbb{Z}~.$
  
\paragraph{$p=2$:}  The $\mathbb{Z}_{2}$ quotient acts simultaneously on the $U(1)$ and $SU(2)$ factors of the group.  All matter fields are neutral under this $\mathbb{Z}_2$ and therefore fractional magnetic charges are allowed with the fractions related as:
    \begin{equation}\label{frac2}
        \frac{F_{1}}{2\pi}=\frac{1}{2}\omega(A_{2})+X~,
    \end{equation}
    where $\omega(A_{2})$ is the discrete magnetic flux of $SU(2)/\mathbb{Z}_{2}$ and $X\in H^{2}(M,\mathbb{Z})$ is an integrally quantized flux.  Making use of \eqref{fracform} which specifies the fractional part of the instanton number, the condition that the axion periodicity $a\sim a +2\pi f$ is respected is thus:
    \begin{equation}\label{2const}
        \frac{\ell_{1}}{2}\int_{M}\left(\frac{\omega(A_{2})}{2}+X\right)^2+\ell_{2}\left(n_{2}+\int_{M}\frac{\omega(A_{2})^2}{4}\right)+\ell_{3}\left(n_{3}\right)\in \mathbb{Z}~,
    \end{equation}
    where above $n_{2}$ and $n_{3}$ denote the integer part of the $SU(2)$ and $SU(3)$ instanton numbers and square is shorthand for wedge/cup product.  Note that this equation must hold for all values of $n_{i}$ and choices of $X$ and $\omega(A_{2}).$  By inspecting the case when $\omega(A_{2})^2$ vanishes (but not the cross terms e.g.~$\omega (A_2) \wedge X$), we deduce that, $\ell_{1} \in 2\mathbb{Z}$ while $\ell_{2}, \ell_{3} \in \mathbb{Z}$.  When this holds, \eqref{2const} reduces to:
    \begin{equation}
        \left(\frac{\ell_{1}}{4}+\frac{\ell_{2}}{2}\right)\int_{M}\frac{\omega(A_{2})^{2}}{2} \in \mathbb{Z}~.
    \end{equation}
    The integral above is always an integer on spin manifolds (where fermions can be defined) and hence we find also: $\ell_{1}+2\ell_{2}\in 4\mathbb{Z}.$

\paragraph{$p=3$:}   The $\mathbb{Z}_{3}$ quotient acts simultaneously on the $U(1)$ and $SU(3)$ factors of the group.  Therefore the fractional magnetic charges are specified by:
    \begin{equation}\label{frac3}
        \frac{F_{1}}{2\pi}=\frac{1}{3}\omega(A_{3})+X~,
    \end{equation}
    where $\omega(A_{3})$ is the discrete magnetic flux of $SU(3)/\mathbb{Z}_{3}$ and $X\in H^{2}(M,\mathbb{Z})$ is an integrally quantized flux.  Again using \eqref{fracform} the condition that the axion periodicity $a\sim a +2\pi f$ is respected is:
    \begin{equation}\label{3const}
        \frac{\ell_{1}}{2}\int_{M}\left(\frac{\omega(A_{3})}{3}+X\right)^{2}+\ell_{2}\left(n_{2}\right)+\ell_{3}\left(n_{3}+\int_{M}\frac{\omega(A_{3})^{2}}{3}\right)\in \mathbb{Z}~,
    \end{equation}
     By inspecting the case when $\omega(A_{3})^{2}$ vanishes, we deduce that, $\ell_{1} \in 3\mathbb{Z}$ while $\ell_{2}, \ell_{3} \in \mathbb{Z}$.  When this holds, \eqref{3const} reduces to:
    \begin{equation}
        \left(\frac{\ell_{1}}{9}+\frac{2\ell_{3}}{3}\right)\int_{M}\frac{\omega(A_{3})^{2}}{2} \in \mathbb{Z}~.
    \end{equation}
Hence we also find: $\ell_{1}+6\ell_{3}\in 9\mathbb{Z}.$

\paragraph{$p=6$:}  Now the quotient acts on all three factors of the gauge group.  Therefore fractional magnetic charges for all three factors of the gauge group are allowed with fractions related as:
\begin{equation}\label{frac6}
     \frac{F_{1}}{2\pi}=\frac{1}{2}\omega(A_{2})+\frac{1}{3}\omega(A_{3})+X~,
\end{equation}
   where  $\omega(A_{2})$ is the discrete magnetic flux of $SU(2)/\mathbb{Z}_{2},$ $\omega(A_{3})$ is the discrete magnetic flux of $SU(3)/\mathbb{Z}_{3},$ and $X\in H^{2}(M,\mathbb{Z})$ is an integrally quantized fluxes.
Using \eqref{fracform}, the condition that the axion periodicity $a\sim a +2\pi f$ is respected is thus:
   \begin{equation}\label{6const}
        \frac{\ell_{1}}{2}\int_{M}\left(\frac{\omega(A_{2})}{2}+\frac{\omega(A_{3})}{3}+X\right)^{2}+\ell_{2}\left(n_{2}+\int_M{}\frac{\omega(A_{2})^{2}}{4}\right)+\ell_{3}\left(n_{3}+\int_{M}\frac{\omega(A_{3})^{2}}{3}\right)\in \mathbb{Z}~,
    \end{equation}
where again $n_{i}$ are the integral instanton numbers.  From the case where both $\omega(A_{2})^{2}$ and $\omega(A_{3})^{2}$ vanish, we deduce that $\ell_{1}\in 6 \mathbb{Z}$, while $\ell_{2}, \ell_{3} \in \mathbb{Z}.$  When this holds, \eqref{6const} reduces to 
\begin{equation}
     \left(\frac{\ell_{1}}{4}+\frac{\ell_{2}}{2}\right)\int_{M}\frac{\omega(A_{2})^{2}}{2}+   \left(\frac{\ell_{1}}{9}+\frac{2\ell_{3}}{3}\right)\int_{M}\frac{\omega(A_{3})^{2}}{2} \in \mathbb{Z}~.
\end{equation}
Therefore we also deduce that $\ell_{1}+2\ell_{2}\in 4\mathbb{Z}$ while $\ell_{1}+6\ell_{3}\in 9\mathbb{Z}.$

The results of this section are summarized in \autoref{tab:SMrules} below.
\renewcommand{\arraystretch}{1.5}
\begin{table}[h!]
\begin{center}
\begin{tabular}{ |llc| }
\multicolumn{3}{c}{Selections rules for $\ell_{1,2,3}$ with $G_{\rm SM} = SU(3)\times SU(2) \times U(1)/\Gamma$} \\
\hline \hline
  $\Gamma=\mathbf{1}$:   & $\ell_{1,2,3} \in \mathbb{Z} $  &  \\
  \hline 
  $\Gamma=\mathbb{Z}_2$:  & $\ell_{1} \in 2\mathbb{Z}$,  $\ell_{2}, \ell_{3} \in \mathbb{Z}$, and $\ell_{1}+2\ell_{2}\in 4\mathbb{Z}$  & \\
  \hline 
 $\Gamma=\mathbb{Z}_3$:  & $\ell_{1} \in 3\mathbb{Z}$,  $\ell_{2}, \ell_{3} \in \mathbb{Z}$, and $\ell_{1}+6\ell_{3}\in 9\mathbb{Z}$ & \\
   \hline 
 $\Gamma=\mathbb{Z}_6$:  & $\ell_{1} \in 6\mathbb{Z}$,  $\ell_{2}, \ell_{3} \in \mathbb{Z}$,  $\ell_{1}+2\ell_{2}\in 4\mathbb{Z}$, and $\ell_{1}+6\ell_{3}\in 9\mathbb{Z}$ & \\
  \hline \hline
\end{tabular}
\caption{\label{tab:SMrules} The selection rules for coefficients $\ell_{1,2,3}$ in the SM, with  $G_{\rm SM} = SU(3)\times SU(2) \times U(1)/\Gamma$, $\Gamma=\mathbf{1}$, $\mathbb{Z}_2$, $\mathbb{Z}_3$, and $\mathbb{Z}_6$. }
\end{center}
\end{table}
\renewcommand{\arraystretch}{1}

\subsection{Axion Potential and Non-Invertible Shift Symmetry}

Next, we turn to the axion potential in the Standard Model coupled to an axion described by the general action \eqref{smact}.

As discussed below \eqref{potp} only integral instantons of the non-abelian factors of the gauge group contribute and hence the potential can be written uniformly independent of the global form of the Standard Model gauge group:
\begin{equation}\label{potsm}
    V(a)=\sum_{n,m=0}^{\infty} \alpha_{n}\beta_{m}\cos\left(\frac{\ell_{2}n a}{f}+\frac{\ell_{3}m a}{f}\right)~,
\end{equation}
where now $\alpha_{n}\propto \exp(-\frac{8\pi^{2}n}{g_{2}^{2}})$ includes the instanton action from the $SU(2)_{L}$ factor of the gauge group, and $\beta_{n}$ includes the instanton action and strong dynamics from the $SU(3)_{C}$ factor of the gauge group.  It is transparent to investigate the symmetries and minima of such a potential.  Indeed, residual shift symmetries arise exactly when the two independent Fourier sums above have a common period which is smaller than $\Delta a =2\pi f.$  This happens precisely when $\gcd(\ell_{2}, \ell_{3})>1,$ which leads to a symmetry: $a\sim a+ \frac{2\pi f}{\gcd(\ell_{2}, \ell_{3})}.$  

While the above analysis is correct, it is crucial to notice the various distinct sizes of the contributions to the axion potential.  The long-distance contributions from $SU(2)_{L}$ can be estimated by evaluating the instanton action at the weak scale.  This gives:
\begin{equation}\label{weaksize}
    \alpha_{1}\sim v^{4}\exp(-\frac{8\pi^{2}}{g_{2}(v)^{2}})\approx (250 \text{GeV})^{4}\exp(-2\pi \cdot 30)\sim (10^{-18}\text{GeV})^4~.
\end{equation}
Meanwhile the analogous contributions from $SU(3)_{C}$ can be estimated from the QCD scale as roughly:
\begin{equation}\label{strongsize}
    \beta_{1}\sim \Lambda_{\text{QCD}}^{4}\approx (10^{-1}\text{GeV})^{4}~.
\end{equation}
The large hierarchy between \eqref{weaksize} and \eqref{strongsize} means that the $SU(2)_{L}$ instantons give very small perturbations to the potential arising from $SU(3)_{C}.$  Hence, up to small corrections we may say that the axion potential \eqref{potsm} has $\ell_{3}$ nearly degenerate minima.  In particular, in the context of the domain wall problem, the contributions to the potential from weak instantons are not large enough to destabilize the nearly topological domain walls arising when $\ell_{3}>1.$  Thus, any such model has a domain wall problem that must be addressed.  

For each of the possible global forms of the Standard Model gauge group, we now determine the nature of the global symmetry of the axion sector in more detail.  In particular, we will deduce when this symmetry is invertible/non-invertible.  To carry this out we note that a symmetry $a\rightarrow a +2\pi f /z$ is invertible if the full action \eqref{smact} is invariant even in the presence of abelian or fractional instantons.  By contrast, when only the potential is invariant, but the action is not, the symmetry is non-invertible. This analysis is nearly identical to that of \autoref{sec:couplings}.

To take into account the hierarchy discussed above, we present both the exact symmetry analysis as well as the approximate symmetry analysis that results from considering contributions to the axion potential from only $SU(3)_{C},$ and neglecting the pieces of \eqref{potsm} that result from  $SU(2)_{L}$ instantons.  Note that even when we neglect $SU(2)_{L}$ contributions to the potential we must still track any potential fractional $SU(2)_{L}$ instanton contributions to correctly determine the nature of the symmetry.  This is because the fusion algebra \eqref{fusion} is an algebraic multiplication rule with quantized coefficients that do not depend on the size of the $SU(2)_{L}$ gauge coupling.

\paragraph{$p=1$:}  Since there is no quotient, there are no fractional instantons.  Denoting by $n_{i}\in \mathbb{Z}$ the integral instanton numbers, the condition that a shift $a\rightarrow a +2\pi f /z$ is an invertible symmetry is:
\begin{equation}
    \frac{\ell_{1}n_{1}}{z}+\frac{\ell_{2}n_{2}}{z}+\frac{\ell_{3}n_{3}}{z}\in \mathbb{Z}~.
\end{equation}
Thus $z$ must divide each of the $\ell_{i}$ and hence divides the greatest common divisor $\gcd(\ell_{1}, \ell_{2}, \ell_{3}).$  The symmetry structure is thus:
\begin{equation}
    \mathbb{Z}_{\gcd(\ell_{1}, \ell_{2}, \ell_{3})}^{I}\subseteq  \mathbb{Z}_{\gcd( \ell_{2}, \ell_{3})}^{NI}\subsetsim  \mathbb{Z}_{\ell_{3}}^{NI}~,
\end{equation}
where the first two factors above indicate the exact invertible and non-invertible symmetry and in the last inclusion, the symbol $\subsetsim$ indicates the approximate symmetry from ignoring $SU(2)_{L}$ instanton contributions to the potential.  Notice that even in this case, where there is no quotient there is still non-invertible symmetry in general.  This is because abelian hypercharge instantons can only be activated in the presence of magnetic charges and hence are effectively all fractional.

\paragraph{$p=2$:}  Now we take into account the fractional magnetic charges specified in \eqref{frac2}.  The condition that a shift $a\rightarrow a +2\pi f /z$ is an invertible symmetry is therefore:
\begin{equation}
     \frac{\ell_{1}}{2z}\int_{M}\left(\frac{\omega(A_{2})}{2}+X\right)^2+\frac{\ell_{2}}{z}\left(n_{2}+\int_{M}\frac{\omega(A_{2})^2}{4}\right)+\frac{\ell_{3}}{z}\left(n_{3}\right)\in \mathbb{Z}~,
\end{equation}
Inspecting the case where $\omega(A_{2})^{2}$ vanishes we deduce that $z$ must divide $\ell_{1}/2, \ell_{2}, \ell_{3}$.  When this is so the condition above reduces to: 
\begin{equation}
    \frac{1}{4z}\left(\ell_{1}+2\ell_{2}\right)\int_{M}\frac{\omega(A_{2})^{2}}{2}\in \mathbb{Z}~.
\end{equation}
Thus $z$ also divides $(\ell_{1}+2\ell_{2})/4$.  We conclude that the symmetry structure is: 
\begin{equation}
    \mathbb{Z}_{\gcd(\ell_{1}/2, \ell_{2}, \ell_{3}, (\ell_{1}+2\ell_{2})/4)}^{I}\subseteq \mathbb{Z}_{\gcd(\ell_{2}, \ell_{3})}^{NI}\subsetsim  \mathbb{Z}_{\ell_{3}}^{NI}~,
\end{equation}
where again the last inclusion is the approximate symmetry from neglecting $SU(2)_{L}$ instanton contributions to the potential.

\paragraph{$p=3$:} The fractional magnetic charges are specified in \eqref{frac3}.  The condition that a shift $a\rightarrow a +2\pi f /z$ is an invertible symmetry is therefore:
\begin{equation}
     \frac{\ell_{1}}{2z}\int_{M}\left(\frac{\omega(A_{3})}{3}+X\right)^2+\frac{\ell_{2}}{z}\left(n_{2}\right)+\frac{\ell_{3}}{z}\left(n_{3}+\int_{M}\frac{\omega(A_{3})^2}{3}\right)\in \mathbb{Z}~,
\end{equation}
Inspecting the case where $\omega(A_{3})^{2}$ vanishes we deduce that $z$ must divide $\ell_{1}/3, \ell_{2}, \ell_{3}$.  When this is so the condition above reduces to:
\begin{equation}
    \frac{1}{9z}\left(\ell_{1}+6\ell_{3}\right)\int_{M}\frac{\omega(A_{3})^{2}}{2}\in \mathbb{Z}~.
\end{equation}
Thus $z$ also divides $(\ell_{1}+6\ell_{3})/9$.  We conclude that the exact symmetry structure is: 
\begin{equation}\label{symform}
    \mathbb{Z}_{\gcd(\ell_{1}/3, \ell_{2}, \ell_{3}, (\ell_{1}+6\ell_{3})/9)}^{I}\subseteq \mathbb{Z}_{\gcd(\ell_{2}, \ell_{3})}^{NI}\subsetsim  \mathbb{Z}_{\ell_{3}}^{NI}~,
\end{equation}
where again the last inclusion is the approximate symmetry from neglecting $SU(2)_{L}$ instanton contributions to the potential.

\paragraph{$p=6$:} Finally, we take into account the fractional magnetic charges in \eqref{frac6}.  The condition that a shift $a\rightarrow a +2\pi f /z$ is an invertible symmetry is therefore:
   \begin{equation}\label{6constz}
        \frac{\ell_{1}}{2z}\int_{M}\left(\frac{\omega(A_{2})}{2}+\frac{\omega(A_{3})}{3}+X\right)^{2}+\frac{\ell_{2}}{z}\left(n_{2}+\int_{M}\frac{\omega(A_{2})^{2}}{4}\right)+\frac{\ell_{3}}{z}\left(n_{3}+\int_{M}\frac{\omega(A_{3})^{2}}{3}\right)\in \mathbb{Z}~.
    \end{equation}
    From the case where $\omega(A_{i})^{2}$ vanishes we learn that $z$ divides $\ell_{1}/6, \ell_{2}, \ell_{3}.$  When this holds we then require
    \begin{equation}
        \frac{1}{4z}\left(\ell_{1}+2\ell_{2}\right)\int_{M}\frac{\omega(A_{2})^2}{2}+   \frac{1}{9z}\left(\ell_{1}+6\ell_{3}\right)\int_{M}\frac{\omega(A_{3})^2}{2} \in \mathbb{Z}~.
    \end{equation}
Therefore, $z$ also divides both $(\ell_{1}+2\ell_{2})/4$ and $(\ell_{1}+6\ell_{3})/9$.  The exact symmetry structure is thus:
\begin{equation}
    \mathbb{Z}_{\gcd(\ell_{1}/6, \ell_{2}, \ell_{3}, (\ell_{1}+2\ell_{2})/4, (\ell_{1}+6\ell_{3})/9)}^{I}\subset \mathbb{Z}_{\gcd(\ell_{2}, \ell_{3})}^{NI}\subsetsim  \mathbb{Z}_{\ell_{3}}^{NI}~,
\end{equation}
where again the last inclusion is the approximate symmetry from neglecting $SU(2)_{L}$ instanton contributions to the potential.

The results of this analysis are summarized in \autoref{tab:SMsym} below. 
\renewcommand{\arraystretch}{1.5}
\begin{table}[h!]
\begin{center}
\begin{tabular}{ |cl| }
\multicolumn{2}{c}{Invertible Axion Shift Symmetry $\mathbb{Z}_{K}^{I}$} \\
\hline \hline
  \hline 
  $\Gamma=\mathbf{1}:$  & $K=\gcd(\ell_{1}, \ell_{2}, \ell_{3})$ \\
  \hline 
   $\Gamma=\mathbb{Z}_{2}:$  & $K=\gcd\left(\frac{\ell_{1}}{2}, \ell_{2}, \ell_{3}, \frac{\ell_{1}+2\ell_{2}}{4}\right)$ \\
  \hline 
   $\Gamma=\mathbb{Z}_{3}:$  & $K=\gcd\left(\frac{\ell_{1}}{3}, \ell_{2}, \ell_{3}, \frac{\ell_{1}+6\ell_{3}}{9}\right)$ \\
  \hline 
   $\Gamma=\mathbb{Z}_{6}:$  & $K=\gcd\left(\frac{\ell_{1}}{6}, \ell_{2}, \ell_{3},\frac{\ell_{1}+2\ell_{2}}{4}, \frac{\ell_{1}+6\ell_{3}}{9}\right)$ \\
   \hline \hline
\end{tabular}
\caption{\label{tab:SMsym} Symmetries of the Standard Model coupled to an axion with couplings $\ell_{i}$.  For each choice of global structure $\Gamma$, we indicate the invertible shift symmetry $\mathbb{Z}_{K}^{I}$ which acts on the axion as $a\rightarrow a+\frac{2\pi f}{K}$.  The associated domain walls are standard and do not support topological degrees of freedom.  The full symmetry structure is $\mathbb{Z}_{K}^{I}\subseteq \mathbb{Z}_{\gcd(\ell_{2},\ell_{3})}^{NI}\subsetsim \mathbb{Z}_{\ell_{3}}^{NI}$, where the last factor is the approximate symmetry that results from neglecting the contributions to the potential from $SU(2)_{L}$ instantons. }
\end{center}
\end{table}
\renewcommand{\arraystretch}{1}

In summary, we see that the typical axion domain wall is a non-invertible topological defect, supporting Chern-Simons gauge fields which couple to the bulk gauge fields through the magnetic one-form symmetry (discrete magnetic fluxes) as in \eqref{dwaction}.   In the ultraviolet, we expect that the non-invertible axion domain walls are lifted and the corresponding non-invertible symmetry is broken.  As described in \autoref{sec:si}, this is suggestive of small instanton effects, or equivalently virtual magnetic monopoles arising from short distance gauge dynamics. In more detail, using the results of \autoref{tab:SMsym}, one may deduce exactly when the fusion of a domain wall $\mathcal{D}$ with its antiwall creates magnetic one-form symmetry defects and hence deduce the precise set of dynamical UV monopoles needed to break both the magnetic one-form symmetry and the non-invertible axion shift symmetry.  In this way the pattern of symmetry deduced in this section can provide a guide towards interesting UV completions.

\subsection{GUT Constraints on Axion Couplings}

In this section, we consider several GUT theories \cite{Georgi:1974sy,Fritzsch:1974nn,Gursey:1975ki}, as well as cases with partial unification \cite{Pati:1974yy,Babu:1985gi}. We derive the formulae of IR anomaly coefficients $\ell_1, \ell_2, \ell_3$ in terms of those appearing in these UV theories. These relationships satisfy all the selection rules on $\ell_1, \ell_2, \ell_3$ shown in \autoref{tab:SMrules}. In addition, they impose \emph{stronger} conditions on $\ell_1, \ell_2, \ell_3$, which provide invaluable information to test short distance fate of the SM when $\ell_1, \ell_2, \ell_3$ are determined from observables of the IR theory. 

Axion coupling measurements can potentially test these predictions \cite{Agrawal:2022lsp}. However, such measurements are only sensitive to relations after the fields are canonically normalized. In fact, for GUTs with simple embeddings such as $SU(5), SO(10), E_6$ \cite{Georgi:1974sy,Fritzsch:1974nn,Gursey:1975ki}, there is a universal prediction for anomaly contribution to the ratio of axion photon and QCD couplings to be $8/3$ \cite{Agrawal:2022lsp}. At the same time, as illustrated in this paper, observables from topological objects such as domain walls and monopoles can be sensitive to the individual $\ell_i$s. 

We also find that the most well-known embeddings can not lift the vacuum degeneracies in the axion potential, including $SU(5)$ ($\Gamma=\mathbb{Z}_6$),  $SO(10)$ ($\Gamma=\mathbb{Z}_6$), Pati-Salam ($\Gamma=\mathbb{Z}_3$), Trinification ($\Gamma=\mathbb{Z}_2$), and $E_6$ ($\Gamma=\mathbb{Z}_6$). Therefore, we find that either GUT theory must have minimal anomaly coefficients or should be supplemented with extra structures to cure the axion domain wall problem. Either way, the axion domain wall problem with the non-trivial global structure imposes severe restrictions on the allowed GUT completion of the SM.

\subsubsection{Georgi-Glashow: $SU(5)$}

The $SU(5)$ GUT \cite{Georgi:1974sy} is broken to $G_{\rm SM} = SU(3)_C \times SU(2)_L \times U(1)_Y / \mathbb{Z}_6$. We start with the UV action 
\begin{equation}
    S_{\rm uv} = \frac{i \ell_{\rm uv}}{8\pi^2} \int \frac{a}{f} \mathrm{tr} \left( g \wedge g \right)
\end{equation}
where $\ell_{\rm uv} \in \mathbb{Z}$ is the UV anomaly coefficient and $g$ is the 2-form field strength of $SU(5)$. 
Since $SU(3)_C \times SU(2)_L$ (algebra level) is embedded in $SU(5)$ as 
\begin{equation}
    \begin{pmatrix}
    \begin{array}{c|c}
    SU(3)_C & \\
    \hline
     & SU(2)_L        
    \end{array}
    \end{pmatrix} \in SU(5)
\end{equation}
one immediately sees that $\ell_{2}$ and $\ell_3$ terms are matched as
\begin{equation}
      S_{\rm uv} = \frac{i \ell_{\rm uv}}{8\pi^2} \int \frac{a}{f} \mathrm{tr} \left( g \wedge g \right)  \supset \frac{i \ell_{\rm uv}}{8\pi^2} \int \frac{a}{f} \mathrm{tr} \left( F_3 \wedge F_3 \right) +  \frac{i \ell_{\rm uv}}{8\pi^2} \int \frac{a}{f} \mathrm{tr} \left( F_2 \wedge F_2 \right),
\end{equation}

which shows that 
\begin{equation}\label{eq:SU5_relation_l23}
    \ell_2 = \ell_{\rm uv}, \;\;\; \ell_3 = \ell_{\rm uv}.
\end{equation}
To figure out the matching condition for $\ell_1$, we note that the $U(1)_Y$ generator $Y$ is embedded in $SU(5)$ as
\begin{equation}
    Y = \alpha
        \begin{pmatrix}
    \begin{array}{ccc|cc}
    -2 &  &  &  &  \\
     & -2 &  &  &  \\
     &  & -2 &  &  \\
    \hline
     &  &  & +3 &  \\
     &  &  &  & +3 \\
    \end{array}
    \end{pmatrix} \in SU(5)
\end{equation}
with the proportionality factor $\alpha$. This factor can be uniquely fixed from the branching rules of matter fields. For now, we can use the following branching rules to fix $\alpha$. 
\begin{equation}
    \Bar{5} = 
    \begin{pmatrix}
        \begin{array}{c}
             d_1^c \\ d_2^c \\ d_3^c \\ e^- \\ - \nu_e      
        \end{array}
    \end{pmatrix}_L \;\;\; \longrightarrow \;\;\; (\Bar{3},1)_{\frac{1}{3}} + (1,\Bar{2})_{-\frac{1}{2}}.
\end{equation}
The first $(\Bar{3},1)_{\frac{1}{3}}$ corresponds to $SU(2)_L$ singlet $d^c$, i.e.~the first three row entries of $\Bar{5}$, while $SU(2)_L$ doublet lepton $(1,\Bar{2})_{-\frac{1}{2}}$ is embedded in the last two row entries of $\Bar{5}$. The subscripts denote $U(1)_Y$ quantum number. Therefore, the SM hypercharge assignments are obtained with $\alpha = -1/6$. The integrally quantized $U(1)_Y$ generator $\hat{Y}$ is then given by $\hat{Y} = 6 Y$. It is in this normalization in which the $U(1)_{Y}$ gauge boson has integral period $\oint F_1 = 2\pi \mathbb{Z}$. The embedding in $SU(5)$ is given by
\begin{equation}
    A^i L^i \supset B \hat{Y}
\end{equation}
where $A^i, L^i$ are the $SU(5)$ gauge field and generator, respectively, and $B$ is hypercharge gauge boson. Therefore, 
\begin{equation}
     S_{\rm uv} = \frac{i \ell_{\rm uv}}{8\pi^2} \int \frac{a}{f} \mathrm{tr} \left( g \wedge g \right)  \supset \frac{i \ell_{\rm uv} \mathrm{tr} (\hat{Y})^2}{8\pi^2} \int \frac{a}{f} \left( F_1 \wedge F_1 \right) .
\end{equation}
Hence, $\ell_1$ in $SU(5)$ GUT is given by
\begin{equation}\label{eq:SU5_relation_l1}
    \ell_1 = 30 \ell_{\rm uv}
\end{equation}
where we used $\mathrm{tr} (\hat{Y})^2 = 30$.
We can verify that \eqref{eq:SU5_relation_l23} and \eqref{eq:SU5_relation_l1} satisfy the conditions for $\Gamma = \mathbb{Z}_6$ (see \autoref{tab:SMrules}).
\begin{equation}
\begin{array}{l}
    \ell_1 = 30 \ell_{\rm uv} \in 6 \mathbb{Z} \;\; \checkmark \\ 
    \ell_1 + 2 \ell_2 = 30 \ell_{\rm uv} + 2 \ell_{\rm uv} = 32 \ell_{\rm uv} \in 4 \mathbb{Z} \;\; \checkmark  \\
    \ell_1 + 6 \ell_3 = 30 \ell_{\rm uv} + 6 \ell_{\rm uv} = 36 \ell_{\rm uv} \in 9 \mathbb{Z} \;\; \checkmark
\end{array} 
\end{equation}
However, we emphasize that the explicit GUT matching conditions provide more stringent conditions, which in principle can be used to rule out or validate each GUT hypothesis.

We end this section by commenting that in the presence of IR axion domain walls ($\ell_{3}>1$), the $SU(5)$ embedding alone does not lift the degeneracy. This is because the index of embedding of $SU(3)_C$ in $SU(5)$ is 1 as shown above, and hence there exist no small instanton configurations to break the axion shift symmetries appearing in the low energy axion coupled to the SM. 

\subsubsection{Pati-Salam: $SU(4)_C \times SU(2)_L \times SU(2)_R$}

The gauge group of the Pati-Salam model is $SU(4)_C \times SU(2)_L \times SU(2)_R$ \cite{Pati:1974yy}, which is broken to the Standard Model with $\Gamma = \mathbb{Z}_3$.\footnote{One can also consider Pati-Salam GUT based on $SU(4)_C \times SU(2)_L \times SU(2)_R/\mathbb{Z}_2$. In this case, one gets $\Gamma = \mathbb{Z}_6$. Our method of analysis can be straightforwardly applied to this case as well.} 
In the UV, we start with 
\begin{equation}
    S_{\rm uv} = \frac{i \ell_4}{8\pi^2} \int \frac{a}{f} \mathrm{tr} \left( g_c \wedge g_c \right) + \frac{i \ell_L}{8\pi^2} \int \frac{a}{f} \mathrm{tr} \left( g_L \wedge g_L \right) + \frac{i \ell_R}{8\pi^2} \int \frac{a}{f} \mathrm{tr} \left( g_R \wedge g_R \right).
\end{equation}
The factor $SU(2)_L$ is just the standard weak gauge group, while $SU(3)_C$ is embedded in $SU(4)_C$ as the upper $3\times3$ block. Hence:
\begin{equation}\label{eq:PS_relation_l23}
    \ell_3 = \ell_4, \;\;\; \ell_2 = \ell_L.
\end{equation}
The coupling $\ell_1$ can be determined via a similar analysis as in the $SU(5)$ case. In particular, $Y$ should be of the form $Y = \alpha \lambda_{\rm B-L} + \beta T_R^3$ where 
\begin{equation}
    \lambda_{\rm B-L} =
    \begin{pmatrix}
    \begin{array}{cccc}
    1/3 &  &    \\
     & 1/3 & &  \\
     &  & 1/3 &  \\
     & & & -1
    \end{array}
    \end{pmatrix} \in SU(4)_C, \;\;\;\;
    %%%
    T_R^3 =
    \begin{pmatrix}
    \begin{array}{cc}
    1/2 &      \\
     & -1/2    \\
    \end{array}
    \end{pmatrix} \in SU(2)_R. 
\end{equation}
The coefficients $\alpha, \beta$ can be fixed from the branching rules 

\begin{equation}
    \begin{array}{l}
       (4,2,1) \;\; \longrightarrow \;\; (3,2)_{1/6} + (1,2)_{-1/2} \\
       %%%%
       (\Bar{4},1,2) \;\; \longrightarrow \;\; (\Bar{3},1)_{1/3} + (\Bar{3},1)_{-2/3} + (1,1)_1 + (1,1)_0.
    \end{array},
\end{equation}
to be
\begin{equation}
    \alpha=\frac{1}{2}, \ \beta = 1.
\end{equation}

From this we get $Y = \frac{1}{2} \lambda_{\rm B-L} + T_R^3$ and the integrally quantized generator is $\hat{Y} = 6 Y$. This result is all we need to determine $\ell_1$.
\begin{equation}
        S_{\rm uv} \supset  \frac{i \ell_L}{8\pi^2} \int \frac{a}{f} \mathrm{tr} \left( g_L \wedge g_L \right) + \frac{i \ell_R}{8\pi^2} \int \frac{a}{f} \mathrm{tr} \left( g_R \wedge g_R \right) \to \frac{i \ell_1}{8\pi^2} \int \frac{a}{f} F_1 \wedge F_1
\end{equation}
with 
\begin{equation}\label{eq:PS_relation_l1}
     \ell_1 =  \ell_L \mathrm{tr (\lambda_{\rm B-L})^2} + \ell_R \mathrm{tr} (T_R^3)^2  = 12 \ell_L + 18 \ell_R.
\end{equation}
\eqref{eq:PS_relation_l23} and \eqref{eq:PS_relation_l1} indeed satisfy (and are stronger
than) the conditions listed in \autoref{tab:SMrules}. 
\begin{equation}
\begin{array}{l}
    \ell_1 = 12 \ell_L + 18 \ell_R \in 3 \mathbb{Z} \;\; \checkmark \\ 
    \ell_1 + 6 \ell_3 = 18 (\ell_4 + \ell_R) \in 9 \mathbb{Z} \;\; \checkmark
\end{array} 
\end{equation}

\subsubsection{Trinification: $SU(3)_C \times SU(3)_L \times SU(3)_R$}

The gauge group of the trinification GUT is $SU(3)_C \times SU(3)_L \times SU(3)_R $ \cite{Babu:1985gi} and is broken to the Standard Model with $\Gamma = \mathbb{Z}_2$.\footnote{In the original proposal, there is also an additional $\mathbb{Z}_3$ which exchanges the 3 gauge factors. This would imply that UV anomaly coefficients are also the same $\ell_c = \ell_L = \ell_R$. We will not make this assumption here. } 
The UV gauge bosons are organized as $24 = (8,1,1) + (1,8,1) + (1,1,8)$ and fermions are packaged into $27 = (1,3,\Bar{3})+ (\Bar{3},1,3) + (3,\Bar{3},1)$. 
In the UV, we start with 
\begin{equation}
    S_{\rm uv} = \frac{i \ell_c}{8\pi^2} \int \frac{a}{f} \mathrm{tr} \left( g_c \wedge g_c \right) + \frac{i \ell_L}{8\pi^2} \int \frac{a}{f} \mathrm{tr} \left( g_L \wedge g_L \right) + \frac{i \ell_R}{8\pi^2} \int \frac{a}{f} \mathrm{tr} \left( g_R \wedge g_R \right).
\end{equation}
Since the first $SU(3)$ is identified with the Standard Model $SU(3)_C$, we have 
\begin{equation}\label{eq:Trini_relation_l3}
    \ell_3 = \ell_c.
\end{equation}

Similarly, the weak $SU(2)_L$ is embedded in $SU(3)_L$ as the upper $2\times 2$ block. Therefore: 
\begin{equation}\label{eq:Trini_relation_l2}
    \ell_2 = \ell_L.
\end{equation}
It remains to determine $\ell_1$. We proceed by first noting that $Y$ must be a linear combination of the three $U(1)$ generators of the UV gauge group that commute with $SU(2)_{L}$ and $SU(3)_{C}$: $Y = \alpha \lambda_L^8 + \beta \lambda_R^3 + \gamma \lambda_R^8$. Here, $\alpha, \beta, \gamma$ are coefficients to be fixed below and the generators are:
\begin{equation}
    \lambda_L^8 =
    \begin{pmatrix}
    \begin{array}{ccc}
    1 &  &    \\
     & 1 &   \\
     &  & -2   \\
    \end{array}
    \end{pmatrix} \in SU(3)_L, \;\;
    %%%
    \lambda_R^3 =
    \begin{pmatrix}
    \begin{array}{ccc}
    1 &  &    \\
     & -1 &   \\
     &  & 0   \\
    \end{array}
    \end{pmatrix} \in SU(3)_R, \;\;
    %%%
    \lambda_R^8 =
    \begin{pmatrix}
    \begin{array}{ccc}
    1 &  &    \\
     & 1 &   \\
     &  & -2   \\
    \end{array}
    \end{pmatrix} \in SU(3)_R.
\end{equation}
The coefficients $\alpha, \beta, \gamma$ are uniquely fixed by the branching rules
%%%%%%%%%
%%%%%
\begin{equation}
    \begin{array}{l}
       (3,\Bar{3},1) \;\; \longrightarrow \;\; (3,2)_{1/6} + (3,1)_{-1/3} \\
       %%%%
       (\Bar{3},1,3) \;\; \longrightarrow \;\; 2 (\Bar{3},1)_{1/3} + (\Bar{3},1)_{-2/3} \\
       %%%%
       (1,3,\Bar{3}) \;\; \longrightarrow \;\; 2 (1,2)_{-1/2} + (1,2)_{1/2} + 2 (1,1)_0 + (1,1)_1 
    \end{array}
\end{equation}
to be
\begin{equation}
    \alpha = - \frac{1}{6}, \ \beta = - \frac{1}{2}, \ {\rm and } \ \gamma = - \frac{1}{6}. 
\end{equation}
The integrally quantized $U(1)_Y$ generator is then given by $\hat{Y} = 6 Y$. Finally, we get
\begin{equation}
    S_{\rm uv} \supset \frac{i \ell_L}{8\pi^2} \int \frac{a}{f} \mathrm{tr} \left( g_L \wedge g_L \right) + \frac{i \ell_R}{8\pi^2} \int \frac{a}{f} \mathrm{tr} \left( g_R \wedge g_R \right) \to \frac{i \ell_1}{8\pi^2} \int \frac{a}{f} F_1 \wedge F_1
\end{equation}
with
\begin{equation}\label{eq:Trini_relation_l1}
    \ell_1 =  \ell_L \mathrm{tr (\lambda_L^8)^2} + \ell_R \mathrm{tr} \left[ (\lambda_R^8)^2 + 9 (\lambda_R^3)^2 \right]  = 6 \ell_L + 24 \ell_R.
\end{equation}
Again, it is quick to verify that \eqref{eq:Trini_relation_l3}, \eqref{eq:Trini_relation_l2} and \eqref{eq:Trini_relation_l1} are consistent with (and stronger than) $\Gamma = \mathbb{Z}_2$ conditions.

\subsection{Lifting SM Axion Domain Walls with Small Instantons}

It is instructive to give an example where the global structure of the SM and UV small instantons may play a role in the domain wall problem for the QCD axion.  Consider the case where $p=3$, so that the Standard Model group is:
\begin{equation}\label{girfinalex}
   G_{\text{IR}}\cong (SU(3)_{C}\times SU(2)_{L}\times U(1)_{Y})/\mathbb{Z}_{3}~.
\end{equation}
We will consider a lift to a UV group:
\begin{equation}
  G_{\text{UV}} \cong SU(2)_{L}\times SU(9)~.
\end{equation}
Note that $SU(2)_{L}$ is a common factor in both the UV and IR and hence does not play any role in the following.  

To explain the group theory, we make use of the example of Section \ref{toy2}.  We note first that $SU(9)$ admits a subgroup $(SU(3)\times SU(3))/\mathbb{Z}_{3}$ with the branching rule stated in \eqref{subranch2}.  We will interpret one of the factors of this product group as color, and a $U(1)$ inside the other factor as hypercharge.  We thus obtain the sequence of branchings:
\begin{equation}\label{subranch}
    \begin{tabular}{ccccc}
       $SU(9)$  & $\rightarrow$ & $(SU(3)\times SU(3))/\mathbb{Z}_{3}$ & $\rightarrow$ & $(SU(3)\times U(1))/\mathbb{Z}_{3}$  \\
       \hline \\
        $\mathbf{9}$  & $\rightarrow$ & $\mathbf{(3,3)}$ &  $\rightarrow$ & $(\mathbf{3},+1)\oplus (\mathbf{3},+1) \oplus (\mathbf{3},-2) $
    \end{tabular}
\end{equation}
Notice that the $U(1)$ charge assignments above are consistent with the $\mathbb{Z}_{3}$ quotient. Furthermore, by computing the indices of embedding for this decomposition we find:
\begin{equation}\label{index9}
  \ell_{1}=18\ell_{\text{UV}}~, \hspace{.2in}  \ell_{3}=3\ell_{\text{UV}}~, 
\end{equation}
where $\ell_{\text{UV}}\in \mathbb{Z}$ is the axion coupling to $SU(9).$  Notice also that $\ell_{1}\in 3\mathbb{Z}$ and $\ell_{1}+6\ell_{3}\in 9\mathbb{Z}$ so this result is consistent with, but stronger than the general constraint for Standard Model gauge group \eqref{girfinalex}.  

We can now apply this result to the domain wall problem.  An IR observer, inspecting the axion potential finds (neglecting $SU(2)_{L}$ instanton corrections), $\ell_{3}=3\ell_{\rm UV}$ minima for the axion and correspondingly stable domain walls.  Relatedly by applying \eqref{symform} we deduce the IR symmetry structure of this class of models is $\mathbb{Z}_{\ell_{\text{UV}}}^{I}\subset \mathbb{Z}_{3\ell_{\text{UV}}}^{NI}$.  In particular, if $\ell_{\text{UV}}=1$, all the axion domain walls in the infrared are all non-invertible and support fractional quantum hall states.

Next, we turn to the ultraviolet.  The non-trivial index of embedding \eqref{index9} implies that small instantons lift some of the degeneracy in the potential and break the non-invertible symmetries from $\mathbb{Z}_{3\ell_{\text{UV}}}^{NI}$ down to its invertible subgroup, $\mathbb{Z}_{\ell_{\text{UV}}}^{I}.$  Thus, for $\ell_{\rm UV}=1$ all domain walls are destabilized by small instantons, and hence the axion domain wall problem is solved in this model by non-invertible symmetry breaking.

\section{Conclusions}
\label{sec:conclusion}

In this work, we have demonstrated that the global structure of the gauge group has a significant impact on axion physics. In particular, the axion couplings to the SM gauge groups can not only provide valuable information about the global properties of the SM but also give highly suggestive indications of the UV completion of the SM. We emphasized that non-invertible symmetry is a very useful tool in this context. In particular, it predicts the existence of degeneracies in the vacua of the axion potential, which can lead to the formation of axion domain walls. The breaking of the non-invertible symmetry by small instanton effects can lift such degeneracies, providing a natural solution to the cosmological axion domain wall problem. We gave a detailed description of this mechanism using an illustrative example. We then applied it to the SM and constructed an explicit example in which the QCD axion domain wall problem can be solved. 

This work should be viewed as a first step to the physics program of using axion as a window for uncovering the global structure of the SM and its UV completion. There are many interesting directions to pursue further. The connection between the details of the gravitational wave signal in connection with the mechanism discussed in the paper needs to be worked out. Additional cosmological and astrophysical observations, such as axion structures formed as a result of the domain walls, may also be relevant as well. As we described in \autoref{sec:theWall}, the non-invertible domain walls (those rooted in the global structure of the gauge group) are different. There are new degrees of freedom on the wall that can potentially be excited when probed by magnetic charges, which can in principle reveal even more information. These interactions between the axion domain wall and magnetic objects are parallel to recent discussions of the Callan-Rubakov effect and generalized symmetries \cite{vanBeest:2023dbu, Brennan:2023tae}, and suggest a more direct connection.  Further investigation of the potential phenomenological consequences of these excitations and interactions will be needed.

\vspace{1cm}
\noindent{\bf Note added: }  Near the completion of this work we received a draft of \cite{Choi:2023pdp} which also studies the quantization conditions on the axion couplings to the Standard Model.  Prior to our submission, \cite{Reece:2023iqn, Agrawal:2023sbp} also appeared with related content.  In particular, \cite{Reece:2023iqn} also derives the quantization conditions on axion couplings for the Standard Model.

\vspace{1cm}
\noindent{\bf Acknowledgements:}
We are grateful to T. D. Brennan, S. Koren, and S. H. Shao for helpful discussions. S. H. would like to thank T. D. Brennan for related collaborations.  We also thank the authors of \cite{Choi:2023pdp} for sharing a draft of their work prior to submission.
We thank the Aspen Center for Physics (supported by a National Science Foundation grant
PHY-2210452) for the opportunity to participate in a summer workshop in 2023, during which part of this work was finished. The work of C.C. is supported by DOE grant DE-SC0024367, by the Simons Collaboration on Global Categorical Symmetries, and by the Sloan Foundation. The work of S.H. is supported by the National Research Foundation of Korea (NRF) Grant RS-2023-00211732. The work of L.T.W. is supported by DOE grant DE-SC-0013642.

\appendix
\section{Cosmology of Long-Lived Axion Domain Walls}
\label{app:axionDWcosmo}

In this section, we briefly summarize the cosmological evolution of the long-lived axion domain walls and the approach of solving the cosmological domain wall problem by introducing a bias in the potential. 
For domain wall number $N>1$, the domain wall string network will be stable. Numerical simulations suggest that the domain walls will be stretched to the horizon size. In this so-called scaling regime, there exists on the order of one domain wall per Hubble patch. Since $H \sim 1/t, \; t=$ the age of the universe, the energy density of the domain walls can be written as
\begin{equation}
\rho_{\rm dw}(t) = d \frac{\sigma}{t},
\end{equation}
where $\sigma$ is the domain wall tension, and $d$ is an ${\mathcal{O}}(1)$ number. For the axion domain wall, we have
\begin{equation}
    \sigma \sim m_a f_a^2. 
\end{equation}
If the domain wall is stable, it will quickly dominate the energy density of the universe, leading to unacceptable cosmology. The time at which the domain wall starts to dominate can be estimated as
\begin{equation}
    \rho_{\rm dw}(t_{\rm dom}) \sim H^2(t_{\rm dom}) M_{\rm Pl}^2, \ \to \ t_{\rm dom} \sim \frac{ M_{\rm Pl}^2}{\sigma} \sim \frac{M_{\rm Pl}^2}{m_a f_a^2}. 
\end{equation}

Lifting the vacuum degeneracy can lead to an imbalance between the pressure on different sides of the domain wall, which could lead to the annihilation of the domain walls. We parameterize the size of splitting in the vacuum energies as 
\begin{equation}
    V_{\rm b} \sim \epsilon f_a^4. 
\end{equation}
Of course, we still have in mind that the global symmetry that leads to the formation of the domain wall is approximately preserved, hence $\epsilon \ll 1$. The annihilation happens when the splitting is comparable to the energy density of the domain wall
\begin{equation}
    \frac{\sigma}{t_{\rm ann}} \sim \epsilon f_a^4, \ \to \ t_{\rm ann} \sim \frac{1}{\epsilon} \frac{m_a}{f_a^2}.
\end{equation}
We would like to have the domain walls annihilate before they dominate the universe
\begin{align}
    t_{\rm ann} < t_{\rm dom}, \ \to \ \epsilon > \left( \frac{m_a}{M_{\rm Pl}}\right)^2
    \label{eq:nodom}
\end{align}

As an example, we consider the QCD axion, with 
\begin{equation}
    m_a = \frac{\Lambda_{\rm QCD}^2}{f_a}, \ \Lambda_{\rm QCD} \sim 0.1 \ \rm{GeV}. 
\end{equation}
Take, for example, $f_a=10^{11}$ GeV. \eqref{eq:nodom} requires  $\epsilon \gtrsim 10^{-62}$.
At the same time, having an extra potential can spoil the axion quality. To avoid that, we require the extra contribution to be much smaller than the size of the QCD generated axion potential $V_{\rm QCD} \sim m_a^2 f_a^2$. That is
\begin{equation}
    \epsilon f_a^4 < \theta_{\rm exp} \times m_a^2 f_a^2,
    \label{eq:strongCP}
\end{equation}
where $\theta_{\rm exp} < 10^{-10}$ is the experimental bound on the strong CP phase. For $f_a=10^{11}$ GeV, \eqref{eq:strongCP} leads to $\epsilon \lesssim  10^{-59}$.

\bibliographystyle{utphys}
\bibliography{AxionDomainWall}

\end{document}